\documentclass[galaxies,article,accept,oneauthor,pdftex,10pt,a4paper]{mdpi}

\firstpage{1}
\articlenumber{64}
\doinum{10.3390/galaxies5040064}
\pubvolume{5(4)}
\pubyear{2017}
\copyrightyear{2017}
\externaleditor{Academic Editors: Markus Boettcher, Emmanouil Angelakis and Jose L. G\'{o}mez}
\history{Received: 15 September 2017; Accepted: 7 October 2017; Published: 12 October 2017}

\pdfoutput=1

\usepackage{bm}
\usepackage{amssymb}
\usepackage{booktabs}
\usepackage{multirow}
\usepackage{soul}
\usepackage{microtype}

\setitemize{parsep=6pt,itemsep=0pt,leftmargin=*,labelsep=5.5mm}
\setenumerate{parsep=6pt,itemsep=0pt,leftmargin=*,labelsep=5.5mm}
\setlist[description]{itemsep=0mm}

\Title{A Model of Polarisation Rotations in Blazars from Kink Instabilities in Relativistic Jets}

\Author{Krzysztof Nalewajko
}

\AuthorNames{Krzysztof Nalewajko}

\address [1] {Nicolaus Copernicus Astronomical Center, Polish Academy of Sciences, Bartycka 18, 00-716 Warsaw, Poland; knalew@camk.edu.pl}

\abstract{This paper presents a simple model of polarisation rotation in optically thin relativistic jets~of~blazars. The model is based on the development of helical (kink) mode of current-driven instability. A possible explanation is suggested for the observational connection between polarisation rotations and~optical/gamma-ray flares in blazars, if the current-driven modes are triggered by secular increases of the total jet power. The importance of intrinsic depolarisation in limiting the amplitude of coherent polarisation rotations is demonstrated. The polarisation rotation amplitude is~thus very~sensitive to the viewing angle, which appears to be inconsistent with the observational estimates of viewing angles in blazars showing polarisation rotations. Overall, there are~serious obstacles to~explaining large-amplitude polarisation rotations in blazars in terms of current-driven kink modes.}
\vspace{-6pt}

\begin{document}

\section{Introduction}

Blazars are cosmic sources of non-thermal radiation produced in relativistic jets~launched by~active galactic nuclei at very small viewing angles. Optical emission of blazars is generally dominated by optically thin synchrotron radiation of relativistic electrons, and as such it shows strong linear polarisation. The apparent polarisation angle is an indirect probe of the orientation of magnetic fields in the main dissipation regions of relativistic jets. Dramatic time variations are observed in blazars, concerning both their total flux, polarised flux and polarisation angle, not necessarily in a correlated fashion. Understanding the complex behaviour of blazars is a long-standing challenge \cite{mad16}.

In recent years, we have seen significant progress in observational monitoring of blazars, including optical, millimeter and radio polarimetry. There were several individual cases of high-energy flares related to an optical polarisation swing and/or passing of a superluminal radio knot through the radio core~\cite{mar08,mar10}. More recently, the connection between gamma-ray flares and optical polarisation rotations in blazars was put on a firm statistical basis by the RoboPol team~\cite{bli15}. It was also noted that blazars can~be~divided into two groups, according to the occurrence of optical polarisation rotations: rotators and~non-rotators~\cite{bli16b}.

Polarisation rotations in blazars have been explained by two classes of models. The first class consists of stochastic models, in which coherent polarisation rotations result from random variations of the polarisation vector~\cite{jon85}.
A significant fraction of polarisation rotations can now~be~explained in~terms~of~a stochastic model~\cite{kie17}.
The second class consists of deterministic models, in~which a~coherent emitting region is assumed to evolve in a way that departs from axial symmetry, e.g.,~propagating on~a~curved trajectory~\cite{nal10,lyu17} or involving regular helical magnetic fields~\cite{zha15}.

In this contribution, the following issues are briefly addressed:
\begin{enumerate}[leftmargin=*,labelsep=5mm]

\item
What could be the mechanism connecting optical polarisation rotations with high-energy flares in~blazars? (Section~\ref{sec_conn})

\item
Is the distinction between rotator and non-rotator blazars due to different viewing angles? (Section~\ref{sec_rot})

\item
The presentation of a novel toy model of polarisation rotations from the development of kink instability in a relativistic conical jet. (Section~\ref{sec_kink})

\item
The importance of intrinsic depolarisation in limiting the amplitude of polarisation rotations.

\end{enumerate}

\section{Results}
\vspace{-6pt}
\subsection{On the Connection between Polarisation Rotations and High-Energy Flares in Blazars}
\label{sec_conn}

Here, by high-energy flares I mean radiation signals produced by the highest-energy particles accelerated in a non-thermal mechanism in the jet dissipation regions. In blazars, high-energy flares are~observed mainly in the gamma-ray band, where they contribute via the inverse Compton process, and~in~the~optical/UV/X-ray band, where they contribute via the synchrotron process. High-energy flares are temporally connected with the radio outbursts and occurrences of superluminal radio knots~\cite{mar12}, the radiation of which is produced by low-energy particles expected to fill a higher fraction of jet volume. The radio emission can be expected to be a better `calorimeter' of~total jet~power, while~the~high-energy flares require localised dissipation regions enabling efficient particle acceleration.

One of the most mysterious observational results on the connection between high-energy and~radio activities of~blazars was a delay between the onset of radio outburst and~the~onset of~gamma-ray flare~\cite{leo11}. This result has been used to suggest that gamma-ray flares are~produced at~large distances, significantly downstream from the radio cores. This would be~in~conflict with the~prevailing scenarios of gamma-ray emission in blazars, in which very dense external radiation fields are~required for~efficient radiative cooling of high-energy electrons~\cite{nal14}. I suggest a possible solution in~which~a~secular increase in the total jet power is a destabilising factor, triggering current-driven kink instabilities, leading to strong magnetic dissipation and efficient particle acceleration. It~has~been recognised that stability of magnetised jets depends crucially on the ratio of toroidal-to-poloidal magnetic field components~\cite{beg98}. Recently, it was realised that the jet environment is~important in~determining the effective magnetic pitch angle, as was illustrated by a distinction between stable headless jets and unstable headed jets~\cite{bro16}. The environmental effect is even stronger if we consider gradual variations in the total jet power~\cite{kom94}. I propose here a preliminary idea that a jet of systematically increasing total power is magnetically overpressured and necessarily belongs to the headed class, and hence increasing jet power could trigger a delayed kink instability, and this in turn can trigger a high-energy flare that is simultaneous with a polarisation rotation.

\subsection{Whether Relativistic Aberration Can Explain the Occurrence of Rotators and Non-Rotators}
\label{sec_rot}

\begin{figure}[H]
\centering
\includegraphics[width=0.9\textwidth]{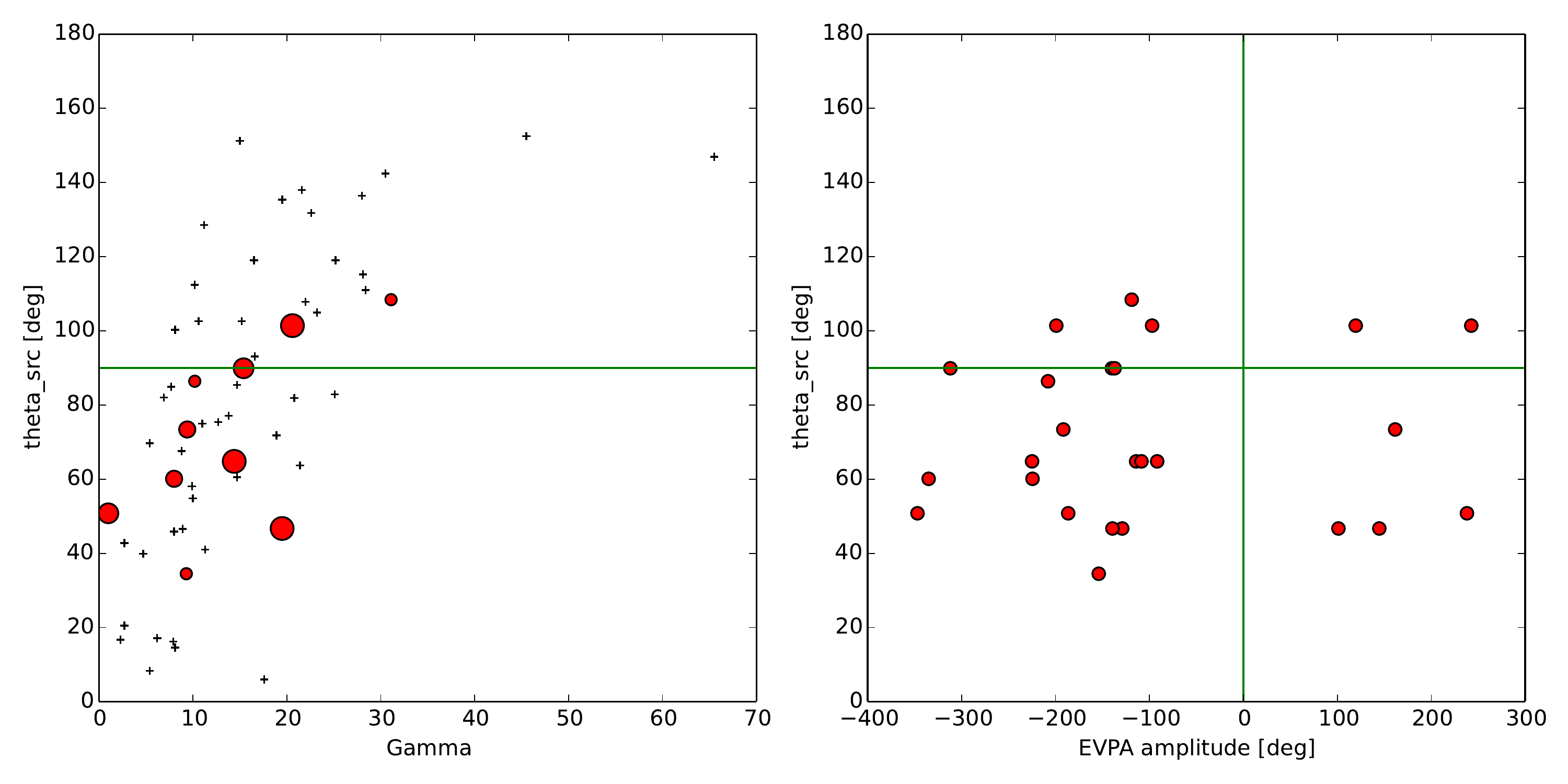}
\caption{\textbf{Left panel}: Intrinsic viewing angle $\theta_{\rm src}$ vs. jet Lorentz factor $\Gamma$ for a sample of bright blazars compiled by~\cite{sav10}. Blazars showing rotations of optical polarisation in the RoboPol survey are indicated with red circles, the size of which indicates the number of rotation episodes; \textbf{Right panel}: $\theta_{\rm src}$ vs. observed polarisation rotation amplitude for individual polarisation rotation episodes.}
\label{fig_theta_src}
\end{figure}

The results of the RoboPol survey indicate that blazars can be divided into two distinct groups: rotators that show (multiple) polarisation rotations and \emph{non-rotators} that do not show any polarisation rotations during the survey~\cite{bli16b}.
The RoboPol team investigated several parameters---blazar class, redshift, Doppler factor, gamma-ray luminosity and variability index---that could potentially reveal systematic differences between rotators and non-rotators, but neither could explain the~occurrence of~polarisation rotations.
To supplement that discussion, I consider here a~possible effect of~the~viewing angle.
It has been possible to estimate the values of intrinsic (co-moving frame) viewing angle $\theta_{\rm src}$ for~a~sample of blazars from independent evaluations of the Lorentz factor $\Gamma = (1-\beta^2)^{-1/2}$ and Doppler factor $\mathcal{D} = \Gamma(1+\beta\theta_{\rm src})$, where $\beta = v/c$ is velocity~\cite{sav10}.
I compare these values for~the~subsamples of rotators and non-rotators. The results are presented in Figure~\ref{fig_theta_src}. I find that the~population of rotators is characterised by $\theta_{\rm src}$ \textasciitilde \,40\textordmasculine--105\textordmasculine, which coincides almost exactly with the range of $\theta_{\rm src}$ occupied by gamma-ray bright blazars~\cite{sav10}.
Moreover, there is no apparent trend between the value of $\theta_{\rm src}$ and the number of rotations observed for individual sources or the observed rotation amplitude.

\subsection{A Model of Polarisation Rotations from Kink Instability}
\label{sec_kink}

Smooth polarisation rotations of amplitudes larger than $180^\circ$, a~good example of~which is~the~rotation observed in source RBPL J1751+0939 reported in~\cite{bli16a}, are the most challenging for~stochastic models and they constitute the strongest case for coherent models. Coherent models often involve compact emitting regions propagating on curved trajectories~\cite{nal10,lyu17}.
However, the~observed emission can also be produced by a pattern propagating in the arbitrary direction with respect to~the~physical flow of the jet.

It is widely acknowledged that current driven instabilities can operate in relativistic magnetised jets, in~which the toroidal component of the magnetic field decays slower than the poloidal component, and~that one~of~the~fastest-growing azimuthal modes is the kink mode ($m = 1$)~\cite{beg98}.
Such a mode is~expected to~introduce a global helical perturbation into the jet structure, which can be described as a perturbation of the radial velocity propagating outwards from the jet core with velocity of the order of the Alfven velocity. I consider a scenario in which this perturbation collides with the jet boundary, triggering dissipation along a helical pattern, intersecting with the roughly conical boundary of the expanding relativistic jet.
A similar scenario in cylindrical geometry, based on relativistic MHD simulations is~described in~\cite{zha17}.

\begin{figure}
\centering
\includegraphics[width=0.9\textwidth]{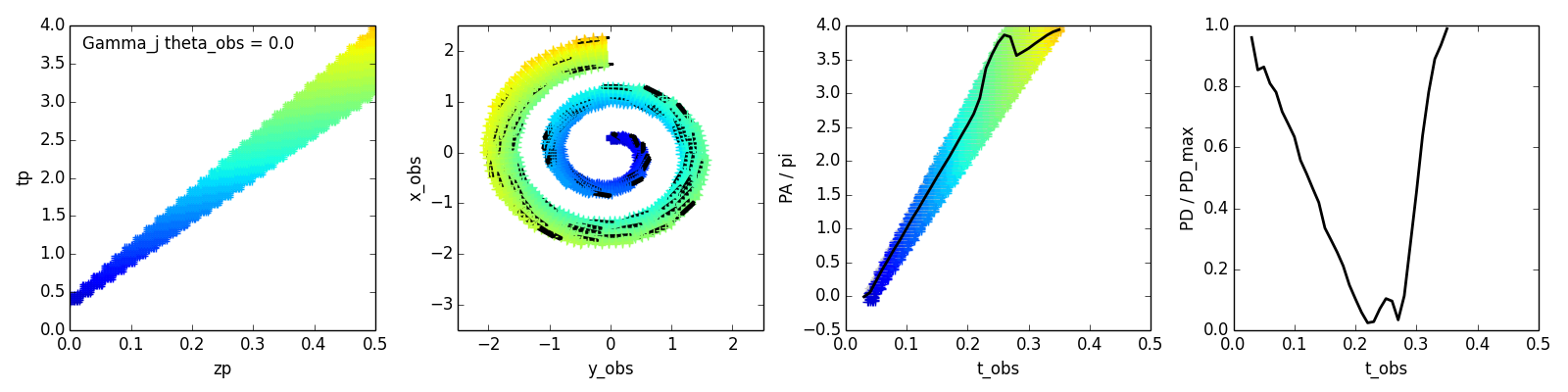}
\includegraphics[width=0.9\textwidth]{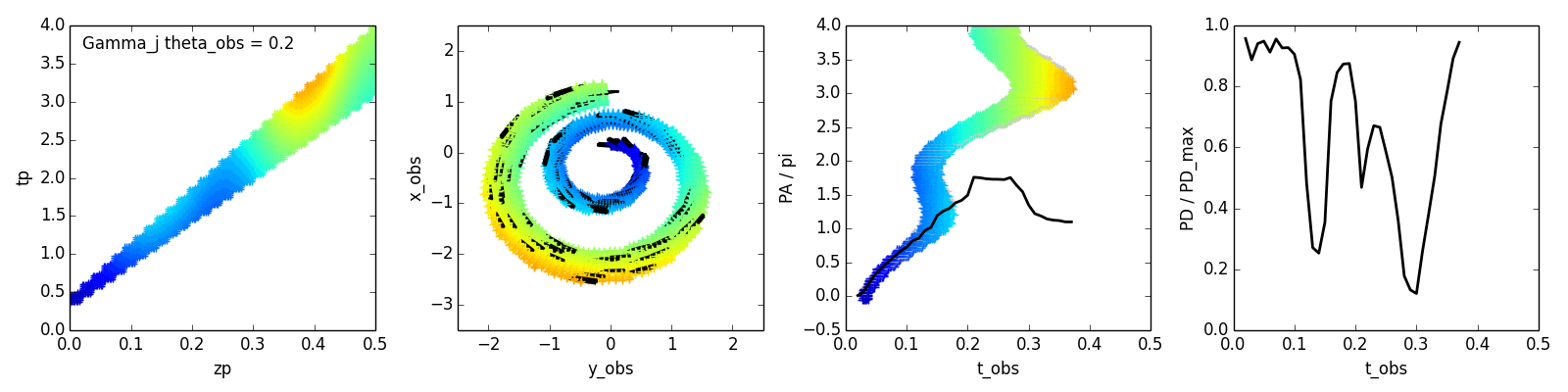}
\includegraphics[width=0.9\textwidth]{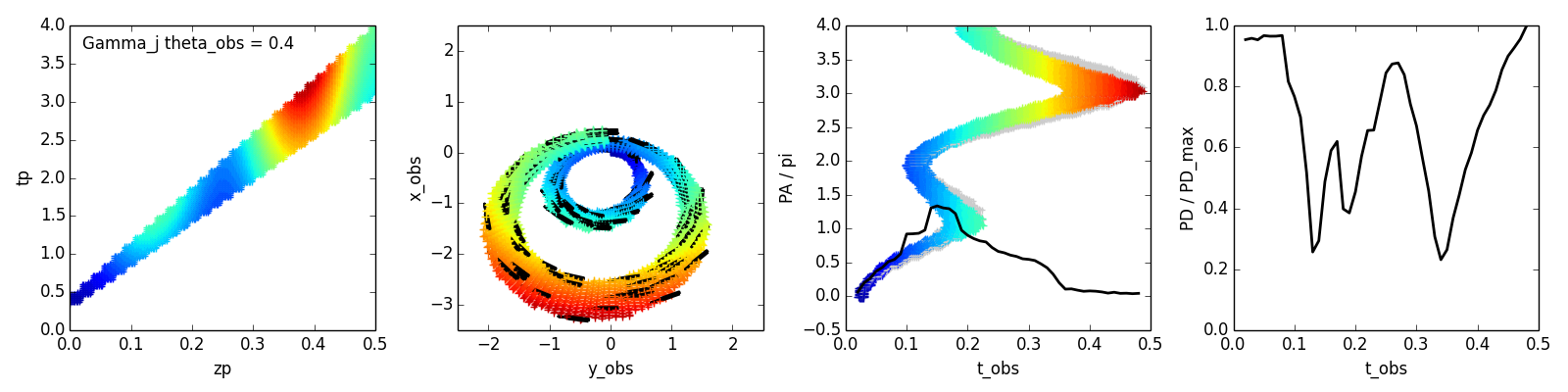}
\caption{Polarisation signal calculated from my model of helical perturbation propagating within a relativistic conical jet for the case of a toroidal magnetic field. The first column shows the emission pattern in co-moving coordinates $t'$ vs. $z'$ with colour indicating the observed time $t_{\rm obs}$. The second column shows the emission pattern in apparent image coordinates $x_{\rm obs}$ vs. $y_{\rm obs}$. The third column shows the observed polarization angle $\chi_{\rm E}$ vs. $t_{\rm obs}$. The black line shows the effective observed signal integrated over $t_{\rm obs}$ bins, and the gray area shows the intrinsic polarization angle (PA) $\chi_{\rm E}'$ vs. $t_{\rm obs}$. The fourth column shows the effective observed polarisation degree (PD) $\Pi$. The rows correspond to different values of the viewing angle $\theta_{\rm obs} = (0,0.2,0.4)/\Gamma_{\rm j}$. Key parameter values are $\Gamma_{\rm j} = 5$, $\Theta_{\rm j} = 0.1$, $\beta_{\rm p} = 0.7$.}
\label{fig_kink_tor}
\end{figure}

\begin{figure}
\centering
\includegraphics[width=0.9\textwidth]{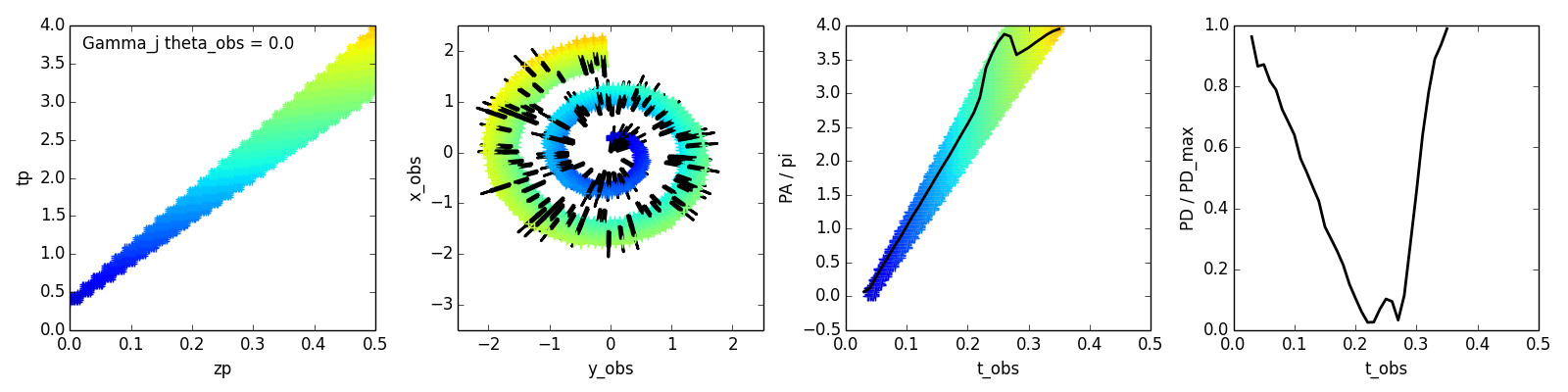}
\includegraphics[width=0.9\textwidth]{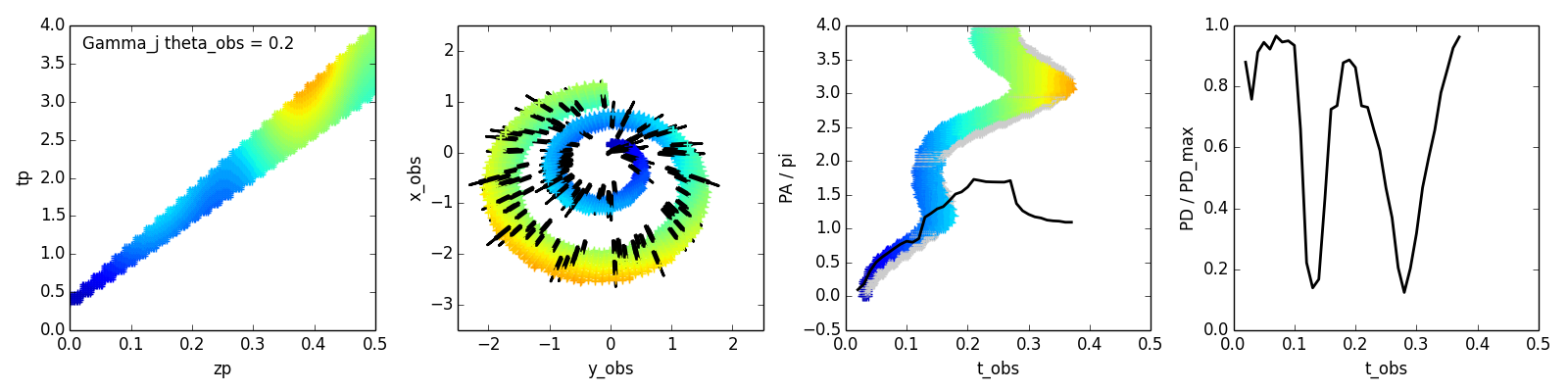}
\includegraphics[width=0.9\textwidth]{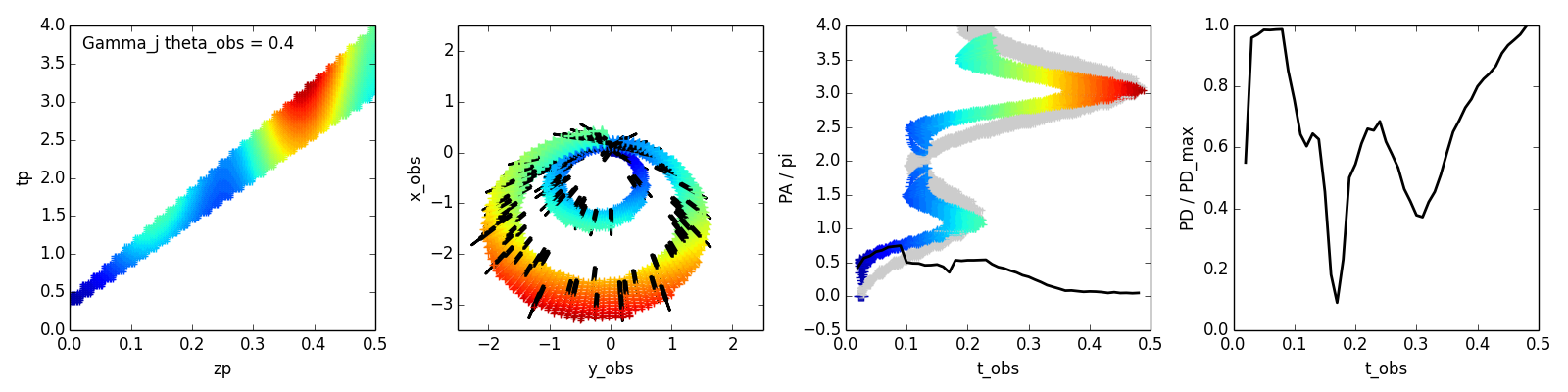}
\caption{Same as Figure~\ref{fig_kink_tor}, but for the case of a poloidal magnetic field.}
\label{fig_kink_pol}
\end{figure}

Details of the model are provided in Section~\ref{sec_methods}.
The results of these calculations are presented in Figure~\ref{fig_kink_tor} for the case of a purely toroidal magnetic field ($B_{\rm pol}' = 0$) and in Figure~\ref{fig_kink_pol} for~the~case of a purely poloidal magnetic field ($B_{\rm tor}' = 0$).
In both figures, I compare the results corresponding to~different viewing angles $\theta_{\rm obs}$.
In the third columns, I show the distributions of polarisation angle $\chi_{\rm E}$ vs. observation time $t_{\rm obs}$.
The intrinsic values of polarisation angle always span the~range of~total kink rotation $\Delta\phi_{\rm tot} = 4\pi$.
However, upon integrating the Stokes parameters, the effective polarisation angle shows a significantly lower amplitude, especially at larger viewing angles.
The fourth columns show effective polarisation degree $\Pi / \Pi_{\rm max}$ as a function of $t_{\rm obs}$.
It is assumed that the emitted radiation is~polarised at~the~fixed level {\bf $\Pi_{\rm max} \sim 70\%$}.
The observed radiation may combine contributions arriving at different polarisation angles, leading to time-dependent depolarisation.
Such depolarisation may~result in~the~departure of the observed polarisation angle from intrinsic values, and~this limits the~observed amplitude of polarisation rotation.
It turns out that such depolarisation is present already for $\theta_{\rm obs} = 0$, due to the finite width of the jet boundary, but it does not yet limit the~amplitude of~polarisation rotation.
However, at larger viewing angles, already at $\theta_{\rm obs} = 0.4/\Gamma_{\rm j}$, the~amplitude of~polarisation rotation is limited to $\Delta\chi_{\rm E} \lesssim \pi$.
The amplitude of polarisation rotation is thus sensitive to~the~viewing angle, but not to the intrinsic orientation of the magnetic field, whether toroidal or poloidal.
{The duration of the observed polarization rotation depends on the distance scale $z_{\rm max}'$ occupied by~the~helical perturbation within the jet. In the example shown in Figures~\ref{fig_kink_tor} and~\ref{fig_kink_pol}, if the spatial units are in pc, polarization rotation would last for $\Delta t_{\rm obs} \sim 0.2\;{\rm pc} \sim 240\;{\rm d}$.}

\section{Materials and Methods: Definition of the Kink Model}
\label{sec_methods}

Let $c = 1$ and $(r,\phi,z)$ be the cylindrical coordinates in which a conical jet of Lorentz factor $\Gamma_{\rm j}$ is~aligned with the $z$ axis, so that $r_{\rm j}=\Theta_{\rm j}z$ is the jet boundary and $\Theta_{\rm j} = 1/\Gamma_{\rm j}$ is the jet opening angle. I~denote the co-moving frame coordinates with a prime. A helical perturbation is~confined to~a~fixed range of co-moving coordinate values $0 < z' < z_{\rm max}'$, with a frozen-in helical pattern defined as~$\phi_{\rm p}(z') = \Delta\phi_{\rm tot}(z'/z_{\rm max}')$, where $\Delta\phi_{\rm tot}$ is the total rotation of the pattern, here set~to~$4\pi$.
At a given distance $z'$, a radial perturbation begins to propagate at a certain moment $t_0'(z')$ according to~$r_{\rm p}'(t',z') = \int_{t_0'}^{t'}{\rm d}t'' \beta_{\rm p}'(t'')$, where $\beta_{\rm p}'(t'-t_0') < 1$ is the co-moving propagation velocity.
In general, one~can~consider various functions for $t_0'(z')$ and $\beta_{\rm p}'(t'-t_0')$.
In the simplest case considered here, I~adopt a~constant propagation velocity $\beta_{\rm p}'$ that applies also to the forward propagation of the perturbation along $z'$, hence $t_0'(z') = z'/\beta_{\rm p}'$ and $r_{\rm p}'(t',z') = \beta_{\rm p}'t'-z'$.
When the perturbation front, Lorentz-transformed to the external frame, reaches the jet boundary defined as $r_{\rm p}/r_{\rm j} \in [0.9:1]$, a polarised emission signal is calculated.
For an observer located at viewing angle $\theta_{\rm obs}$ from the jet axis, i.e., along the unit vector $\bm{k} = [\sin\theta_{\rm obs},0,\cos\theta_{\rm obs}]$, the signal is observed at $t_{\rm obs} = t - k_zz - k_xr_{\rm p}\cos\phi_{\rm p}$.
I use another unit vector along the local direction of the conical jet flow, $\bm{e} = [r_{\rm p}\cos\phi_{\rm p},r_{\rm p}\sin\phi_{\rm p},z]/(r_{\rm p}^2+z^2)^{1/2}$, to~calculate relativistic aberration of the emitted photons~\cite{nal09}: $\bm{k}' = \mathcal{D}[\bm{k} + ((\Gamma_{\rm j}-1)\mu_{\rm obs} - \Gamma_{\rm j}\beta_{\rm j})\bm{e}]$,
where $\mathcal{D} = [\Gamma_{\rm j}(1-\beta_{\rm j}\mu_{\rm obs})]^{-1}$ is the local Doppler factor,
and $\mu_{\rm obs} = \bm{e}\cdot\bm{k}$.
I also select two orthonormal vectors forming a basis in the plane orthogonal to $\bm{k}'$: $\bm{v}' = [k_z',0,-k_x']$ and $\bm{w}' = [-k_x'k_y',1-k_y'^2,-k_y'k_z']$. I now consider magnetic field in the co-moving frame that consists of two components: toroidal $B_{\rm tor}'$ and poloidal $B_{\rm pol}'$, such that $\bm{B}' = [-B_{\rm tor}'\sin\phi_{\rm p},B_{\rm tor}'\cos\phi_{\rm p},B_{\rm pol}']$. I calculate the co-moving magnetic vector position angle $\chi_{\rm B}' = \arctan[(\bm{B}'\cdot\bm{w}')/(\bm{B}'\cdot\bm{v}')]$, the co-moving fluid velocity position angle $\chi_{\rm e}'~=~\arctan[(\bm{e}\cdot\bm{w}')/(\bm{e}\cdot\bm{v}')]$, and the external fluid velocity position angle $\chi_{\rm e} = \arctan[e_y/(e_xk_z-e_zk_x)]$. The observed electric vector position angle is given by $\chi_{\rm E} = \chi_{\rm B}' - \chi_{\rm e}' + \chi_{\rm e} - \pi/2$. Finally, I~calculate the Stokes parameters $\delta I = (\mathcal{D}^3/\Gamma_{\rm j})\delta I'$, $\delta Q = \Pi_{\rm max}\delta I\cos(2\chi_{\rm E})$, and $\delta U = \Pi_{\rm max}\delta I\sin(2\chi_{\rm E})$ that are~eventually integrated in bins of $t_{\rm obs}$, and the observed polarisation angle is calculated taking into account the $\pm 180^\circ$ ambiguity expected for finite time bins.

\section{Conclusions}

 A novel scenario was considered for the production of large-amplitude optical polarisation rotations in~relativistic jets of blazars, that is motivated by the development of current-driven kink instabilities with synchrotron emission confined to the conical boundary layer.
It was found that depolarisation of~radiation signals produced at different jet regions but observed simultaneously has a very strong effect on~the~observed amplitude of polarisation rotation, which is very sensitive to the viewing angle.
My simple model predicts that polarisation rotation amplitude should be limited to $\Delta\chi_{\rm E} \lesssim \pi$ for~$\theta_{\rm obs} \gtrsim 0.4/\Gamma_{\rm j}$.
On the other hand, estimates of $\theta_{\rm src}$ for rotator-class blazars suggest that polarisation rotations of~amplitude $\Delta\chi_{\rm E} > \pi$ are observed solely for $\theta_{\rm obs} \sim (\theta_{\rm src}/90^\circ)/\Gamma_{\rm j} \gtrsim 0.4/\Gamma_{\rm j}$, which coincides exactly with that selecting gamma-ray bright sources~\cite{sav10}.
To conclude, if the values of $\theta_{\rm src}$ estimated by~\cite{sav10} are correct, the viewing angle is unlikely to explain the distinction between rotator and non-rotator blazars.

\vspace{6pt}

\acknowledgments{This work was supported by the Polish National Science Centre grant 2015/18/E/ST9/00580.
}

\conflictsofinterest{The author declares no conflict of interest.
}


\reftitle{References}

\end{document}